\documentclass{article}

\usepackage{arxiv}

\usepackage[utf8]{inputenc} % allow utf-8 input
\usepackage[T1]{fontenc}    % use 8-bit T1 fonts
\usepackage{hyperref}       % hyperlinks
\usepackage{url}            % simple URL typesetting
\usepackage{booktabs}       % professional-quality tables
\usepackage{amsfonts}       % blackboard math symbols
\usepackage{nicefrac}       % compact symbols for 1/2, etc.
\usepackage{microtype}      % microtypography
\usepackage{lipsum}
\usepackage{graphicx}
\usepackage{xcolor}

\title{Exchanging... Watch out!}

\author{Liu Yang \\
	ISIR - Sorbonne University \\
	France \\
	\texttt{yangl@isir.upmc.fr} \\
	\And
	Jieyeon Woo \\
	ISIR - Sorbonne University \\
	France \\
	\texttt{woo@isir.upmc.fr} \\
	\And
	Catherine Achard \\
	ISIR - Sorbonne University \\
	France \\
	\texttt{achard@isir.upmc.fr} \\
        \And
	Catherine Pelachaud \\
	CNRS - ISIR - Sorbonne University \\
	France \\
	\texttt{pelachaud@isir.upmc.fr} \\
}

\begin{document}

\definecolor{darkgreen}{rgb}{0.0, 0.7, 0.0}
\definecolor{lightred}{rgb}{1., 0.44, 0.37}

\maketitle

\begin{abstract}
During a conversation, individuals take turns speaking and engage in exchanges, which can occur smoothly or involve interruptions. Listeners have various ways of participating, such as displaying backchannels, signalling the aim to take a turn, waiting for the speaker to yield the floor, or even interrupting and taking over the conversation.

These exchanges are commonplace in natural interactions. To create realistic and engaging interactions between human participants and embodied conversational agents (ECAs), it is crucial to equip virtual agents with the ability to manage these exchanges. This includes being able to initiate or respond to signals from the human user. In order to achieve this, we annotate, analyze and characterize these exchanges in human-human conversations. In this paper, we present an analysis of multimodal features, with a focus on prosodic features such as pitch (F0) and loudness, as well as facial expressions, to describe different types of exchanges.
\end{abstract}


\maketitle

\section{Introduction}

Human-computer interfaces are becoming increasingly prevalent and valued in everyday life and the development of Embodied Conversational Agents (ECAs) is thriving due to their ability to facilitate natural interactions devoid of artificiality. However, the complexity of natural interactions poses numerous challenges that span multiple research fields, ranging from psychology to signal processing. Extensive work has been conducted on verbal and non-verbal signals, leading to the development of several ECAs. One essential topic for ECA development is to manage the speaking floor, many works exist already on the research of turn-takings and backchannels, while one crucial aspect that remains inadequately explored is interruptions. Interruptions are, however, quite common in natural conversations \cite{beattie1981interruption} and occur when one participant attempts to interject while the other is still speaking. Interruptions are integral to the turn-taking process. In earlier studies, interruptions were often associated with dominance and power dynamics \cite{ferguson1977simultaneous,tannen1991you,itakura2001describing} as most conversations adhere to the rule of one person speaking at a time. However, interruptions play a vital role in natural interactions, helping to regulate dialogue rhythm, express interest and reinforce engagement \cite{yang2001visualizing}.

During natural interactions, speakers exchange speaking floors seamlessly and rapidly. Humans possess the ability to anticipate the conclusion of their partner's turn, allowing them to smoothly take their own turn without disrupting the flow of the conversation \cite{de2006projecting}. Likewise, humans can readily recognize when their partners display backchannels, indicating active participation in the discussion. When an interruption occurs, the current speaker decides whether to yield the floor to the interrupting party.

In order to develop an Embodied Conversational Agent (ECA) capable of engaging in natural interactions with human interlocutors, we believe it is crucial to endow ECAs with the ability to manage different kinds of exchanges such as smooth turn exchanges, backchannels and interruptions \cite{cafaro2016effects}, either by initiating the exchanges or reacting to ones from the human partner. To achieve this, the ECA must discern multimodal signals from its human interlocutor. Additionally, the ECA should be able to identify different types of exchanges. ECAs play not only the role of the speaker, but they also need to initiate exchanges like taking the speaking floor, sending backchannels both verbally and non-verbally, and sometimes interrupting human partners in a meaningful way without being perceived as a system error. To well control ECAs’ behaviours and manage these kinds of exchanges, it is important to learn firstly how humans deal with similar situations, and then model and adjust their behaviours based on the study to make them more lively and realistic.

To accomplish this objective, we analyze natural turn exchanges in human-human interactions using the dyadic corpus NoXi \cite{cafaro2017noxi}. We propose an annotation schema for interruptions and conduct an analysis of multimodal features to examine the non-verbal behaviours associated with each type of turn transition. Our aim is to determine which features humans employ to comprehend these situations and subsequently endow ECAs with similar capabilities. The multimodal features we consider include prosodic features (e.g., pitch and loudness) and facial expressions.

In the upcoming section, we begin by introducing studies concerning turn-taking and, specifically, interruptions in human-human interaction. Moving on to Section 2, we provide an overview of existing researches on classifying turn-taking exchanges, backchannels and interruptions. The subsequent sections cover the NoXi corpus (Section 3) and the statistical outcomes derived from our annotations (Section 4). In Section 5, we present the multimodal features that we have automatically extracted and their analysis.

\section{Related Works}

The concept of interaction has captivated numerous scholars for a considerable period of time. Emanuel A. Schegloff \cite{schegloff1968sequencing} introduced sequencing rules that govern natural conversations. Building upon this, Harvey Sacks \cite{sacks1978simplest} proposed the notion of conversation analysis, highlighting turn-taking as its fundamental structure. In essence, participants in a conversation coordinate and exchange the speaking floor based on established rules, operating under the assumption that speaking and listening cannot occur simultaneously. This fundamental structure of turn-taking was further developed by Kendon \cite{kendon1967some} and Duncan \cite{duncan1972some}, who presented a conversation model incorporating three key signals:
\begin{itemize}
\item Turn-yielding signals from the speaker: These signals indicate that the listener may take the next turn and the speaker yields the floor when the listener displays a willingness to speak.

\item Attempt-suppressing signals from the speaker: These signals are used by the speaker to maintain their turn and prevent the listener from taking over.

\item Backchannel signals from the listener: These signals provide feedback to the speaker but do not signify an attempt to take the turn. They are not considered as a turn themselves.
\end{itemize}
 Sacks, Schegloff and Jefferson \cite{sacks1978simplest} proposed a conversation turn-taking model known as the SSJ model, which outlines the mechanisms involved in turn-taking. The model is based on the following rules:

(\textit{i}) The current speaker may choose the next speaker, who is then obliged to speak.

(\textit{ii}) If the current speaker does not select anyone, a participant may self-select to speak next.

(\textit{iii}) If no one self-selects, the current speaker may choose to continue speaking or terminate the conversation.

Sacks and his colleagues hypothesized that conversational participants accurately predict the timing of turn endings, resulting in a "no gap, no overlap" pattern between turns. However, \cite{coates199411} analyzed the timing intervals during turn exchanges and discovered a significant number of overlapping occurrences at the end of turns in various conversation settings. This finding refuted the hypothesis of "no gap, no overlap."

Schegloff and Sacks \cite{schegloff1973opening} introduced the study of specific speaking turn exchanges that involve simultaneous speeches. These exchanges are categorized as interruption, overlap, or parenthetical comments, such as backchannels. Backchannels are brief messages that indicate the listener's attention or agreement/disagreement with the speaker's speech \cite{allwood1992semantics}.

Overlap occurs when the listener begins speaking before the current speaker has completed their speech, typically overlapping with the last word(s) or syllable(s) of the current speaker and the first word(s) of the listener \cite{sacks1978simplest}. In contrast, interruption happens when the listener forcefully takes the floor against the speaker's will, not allowing them to finish their utterance. Interruption is considered a violation of the current speaker's turn, whereas overlap is not \cite{schegloff1973opening, moerman2010appendix}.

The realm of research emphasizes the significance of prosodic features, such as fundamental frequency (F0) and intensity, during speaking turn exchanges \cite{french1983turn,kurtic2013resources,truong2013classification}. Intriguingly, studies have found that when individuals attempt to interrupt the current speaker, they raise their energy and voice \cite{shriberg2001observations,gravano2012corpus}. Hammarberg's  \cite{hammarberg1980perceptual} work aligns harmoniously with these findings, offering parallel evidence for pitch and amplitude.

Delving deeper, conversational analysts explore an array of characteristics exhibited by interrupters, including speech rate, cutoffs and repetitions. As an example, Schegloff's \cite{schegloff2000overlapping} enlightening discoveries uncover the variations of prosodic profiles and repetitions employed by interrupters. Furthermore, he observes that interrupting sentences often exhibit a faster speaking rate, providing further evidence of the role of speech rate in managing speaker conflicts. 

Gravano and colleagues \cite{gravano2012corpus} conducted an analysis of acoustic features in a telephonic conversation corpus, revealing significant differences in intensity and pitch levels, speaking rate and the duration of Inter-Pausal Units (IPUs) during interruptions. An IPU corresponds to a sequence of words surrounded by silences of $50 ms$ or more.

In the quest to anticipate the conclusion of a speech turn, Riest \textit{et al.} \cite{riest2015anticipation} and De Ruiter \textit{et al.} \cite{de2006projecting} argue in favour of the utility of semantic information, while Van Turnhout \textit{et al.} \cite{van2005anticipating} and Bogels \textit{et al.} \cite{bogels2021turn} highlight the significance of contextual cues. Auer \textit{et al.} \cite{auer2021turn} reported the importance of gaze on the next-speaker selection. According to Zima \textit{et al.}, prevailing speakers, who successfully assert their dominance in the competition for the floor, tend to redirect their gaze away from the competing speaker as a strategic means of maintaining their turn. In addition to gaze direction, Kendrick \textit{et al.} \cite{kendrick2023turn} further mentioned that manual gestures also aid the coordination of turn transitions, the transition of turns is multimodal in nature.

Stivers \textit{et al.} \cite{stivers2009universals} note that, on average, the intervals between speech turns last approximately $200 ms$. However, psycholinguistic research reveals that even a single-word utterance requires a minimum of $600 ms$ to be produced \cite{indefrey2004spatial}. Hence, it becomes evident that the next speaker must engage in certain cognitive processes to predict the termination of the current speaker's turn  \cite{dediu2013antiquity,holler2016turn}. Furthermore, research indicates that predictive mechanisms operate alongside other processing layers, enabling simultaneous planning of production and comprehension \cite{de2006projecting}. Nevertheless, the accuracy of turn-ending timing prediction is occasionally compromised, resulting in instances of simultaneous speeches.

 In this research, we focused on delving into multimodal features to gain a deeper understanding of the nonverbal signals displayed by interlocutors during exchanges. In addition to incorporating acoustic features, hand gestures and gaze, as explored in previous works, we also consider the crucial role of facial expressions. Recognizing the importance of studying the interplay and mutual influence between the interlocutors across different signals and exchange types, we examine the behaviour of both sides of the interaction, rather than solely focusing on the behaviour of either the speaker (the one giving the turn) or the exchange initiator (the one taking the turn).

\section{Corpus}

For our research, we have chosen to utilize the NoXi corpus \cite{cafaro2017noxi} as the foundation of our study. The NoXi corpus is a comprehensive collection of multimodal data, consisting of video and audio recordings capturing natural dyadic interactions. Each interaction involves two participants who have been recorded separately, allowing easy separation of the audio sources. The setup of screen-mediated interactions facilitates clear visual separation. The video recordings capture almost the entire body of each participant, except for their feet. Both the audio and video recordings of the interactants have been synchronized and transcribed.

Within the NoXi database, participants assume either the role of an "expert" or a "novice." The expert imparts their knowledge on a specific subject, selected from a pool of over $45$ given topics to the novice who is interested in that particular subject. Each interaction lasts approximately $20$ minutes, resulting in a rich dataset of diverse and informative interactions.

The NoXi corpus encompasses data recorded in seven different languages. For our study, we specifically selected the French portion of the NoXi corpus, which includes $21$ dyadic conversations, and approximately $7$ hours of recorded interactions in total ($21$ * $20$ minutes).

\section{Annotation \& Statistical Results}

Based on the schema outlined in \cite{anonymous_schema}, exchanges are categorized into three main elements: smooth turn exchange, backchannel, and interruption. Specifically, interruptions are divided into successful and failed interruptions and then into two main types regarding their context:
\begin{itemize}
\item Cooperative: Agreement, clarification, and assistance
\item Competitive: Disagreement, floor taking, topic change and tangentialization
\end{itemize}

We have identified $3983$ voice activity detection (VAD) switch points within the French segment of the NoXi database. Among these points, $1403$ instances correspond to smooth turn exchanges, $1651$ are categorized as backchannels and $929$ are interruptions. %When we exclude backchannels that do not align with speaking turn exchanges, 
Interruptions account for $33 \%$ of turn-taking situations (when floor transition occurs, backchannels excluded). This further supports our belief that interruptions play a crucial role in natural interactions.

Focusing specifically on interruptions, a significant majority of them ($81.7 \%$) are executed successfully, resulting in the interrupting speaker gaining the floor. Interestingly, among these successful interruptions, there is an almost equal distribution between cooperative interruptions ($54.36 \%$) and competitive interruptions.

We examine how long the first IPU lasts after an exchange. smooth turn exchanges have the longest first IPU, with a mean of $4.31$ seconds. Interruptions have a shorter first IPU, with a mean of $2.91$ seconds. The first IPU is longer for successful interruptions (mean = $3.33$ seconds) than for failed ones (mean = $1.04$ seconds). The type of interruption also affects the first IPU length. Competitive interruptions result in longer first IPU (mean = $4.01$ seconds) than cooperative ones (mean = $2.46$ seconds). This is especially true for successful competitive interruptions (mean = $4.47$ seconds), which are $1.89$ seconds longer than successful cooperative ones (mean = $2.58$ seconds). 

\begin{figure}[h]
  \centering
  \includegraphics[width=0.7\linewidth]{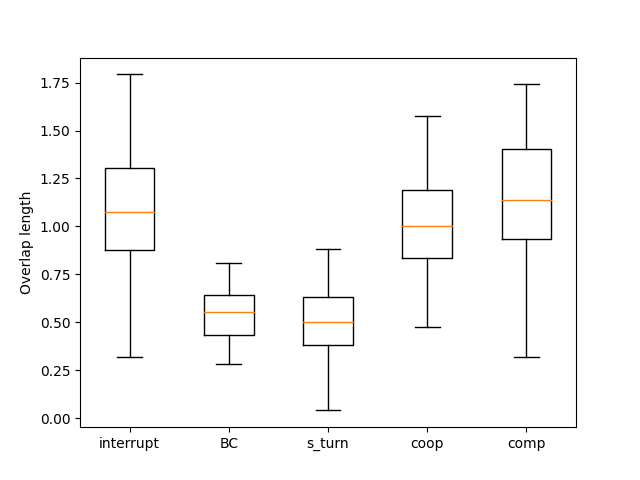}
  \caption{Overlap length for different types of exchanges. Labels: Interruption (interrupt), backchannel (BC), smooth turn exchange (s\_turn), cooperative interruption (coop), competitive interruption (comp).} 
  \label{fig:overlap}
\end{figure}

We observe that overlaps occur frequently in interruptions ($88 \%$) and backchannels ($71 \%$), but rarely in smooth turn exchanges ($29 \%$), where the turn-taker tends to wait until the speaker completes the turn. Figure \ref{fig:overlap} illustrates the variation in overlap duration across exchange types. Interruptions have the longest overlaps, with a mean of $1.15 s$, indicating a high degree of competition for the floor. Smooth turn exchanges have much shorter overlaps than interruptions, with a mean of $0.62 s$, suggesting a low level of conflict and a high level of coordination. Backchannels have the shortest overlap duration, as they are mostly single words or syllables that signal agreement or attention.

\begin{figure}[h]
  \centering
  \includegraphics[width=0.7\linewidth]{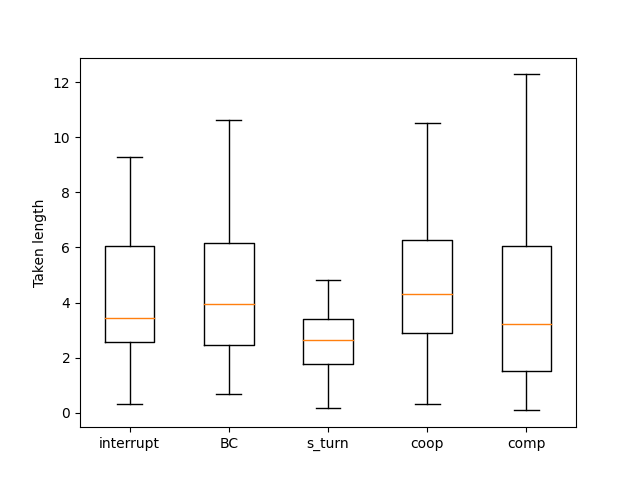}
  \caption{Relative distance between the exchange onset point and the start of the speaker’s last IPU for different types of exchanges. Labels: Interruption (interrupt), backchannel (BC), smooth turn exchange (s\_turn), cooperative interruption (coop), competitive interruption (comp)} 
  \label{fig:taken}
\end{figure}

We also examine the timing of the exchange onset point relative to the start of the speaker’s last IPU, as shown in Figure \ref{fig:taken}. We find that smooth turn exchanges occur earlier after the start of the IPU, with a mean of $2.95 s$, than interruption and backchannels. This may indicate that the speaker gives cues for the turn ending. Interruptions and backchannels happen later, with a mean of $4.38 s$ and $4.93 s$ respectively. This may suggest that the speaker does not intend to yield the floor or the turn-taker needs more time to make the decision. Cooperative interruptions wait a bit longer than backchannels $4.99 s$ on average which has the longest delay, as they are often used to show continued interest or understanding. 

We also investigate the influence of roles (expert, novice) on the conversation dynamics. We find that novice speaks for $29.4 \%$ of the conversation duration on average, while expert speaks for $69.8 \%$ of the conversation duration on average. This reflects the asymmetry of knowledge and authority between the roles. However, we also find that most of the interruptions ($60.6 \%$) and backchannels ($76.5 \%$) are initiated by the novices. This may indicate that the novices are more active and engaged in the conversation, or that they are more likely to challenge or seek clarification from the expert. Furthermore, we measure the length of the first IPU after the exchange and we find that the expert’s first IPU is significantly longer than the novice’s first IPU, both for interruptions ($3.98 s$ vs $2.86 s$) and smooth turn exchanges ($5.49 s$ vs $3.65 s$). This may suggest that the expert has more information to convey or more confidence to hold the floor than the novice.

\section{Multimodal Analysis}

In this section, we describe the multimodal features that we extracted from the data and the methods that we used to segment the data into different types of exchanges. Then, we report the results of our analysis and discuss the main findings.

\subsection{Features \& Segmentation}
We used openSmile\cite{eyben2010opensmile} to extract acoustic features related to pitch (F0) and loudness from the speech signals. We also used OpenFace\cite{baltrusaitis2018openface} to extract facial expressions, gaze, and head movements from the video recordings. These features included Action Units (AUs: AU01(inner brow raiser), AU02(outer brow raiser), AU04(brow lower), AU06(cheek raiser) and AU12(lip corner puller)), which indicate different facial muscle activations. We normalized all these features by conversation and by speaker.

From the extracted AU features (notably AU01, AU02, AU04, and AU12), we obtained high-level synchrony features, along with the low-level signal (acoustic and visual) features, via synchrony measures.
We employed the synchrony measures presented in \cite{anonymous_sync} which have shown the relation between synchrony measures and exchange types (interruption, smooth turn exchange and backchannel). The measures are as follows.
\begin{itemize}
    \item Pearson correlation coefficient (PCC): calculates the association strength between two continuous sequences.
    \item Time-lagged cross-correlation (TLCC): computes the relation between two continuous sequences being invariant to temporal shifts. We chose the time lag of $4$ seconds~\cite{chartrand1999chameleon,leander2012you}.
    \item Dynamic Time Warping (DTW)~\cite{muller2007dynamic}: measures the similarity between two continuous sequences being invariant to length and speed.
\end{itemize}

\begin{figure}[h]
  \centering
  \includegraphics[width=0.7\linewidth]{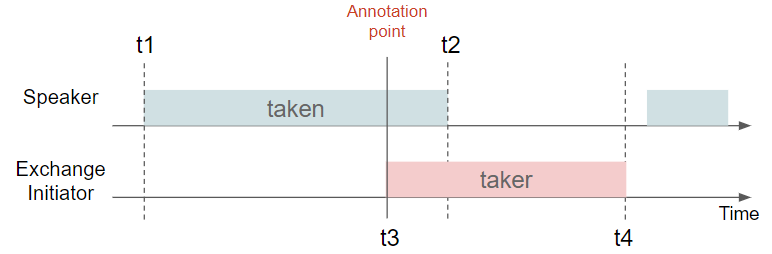}
  \caption{Explanation of data segmentation structure (taken: the last IPU of the speaker before the exchange, taker: the first IPU of the exchange initiator after the exchange onset point).} 
  \label{fig:segment}
\end{figure}

We analyzed these features in the following time intervals: from the start of the last speaker’s IPU (t1) to the end of the last speaker’s IPU (t2), from the end of the last speaker’s IPU (t2) to the start of the first exchange initiator’s IPU (t3, known as the annotation point) and from the start of the last speaker’s IPU (t1) to the end of the first exchange initiator’s IPU (t4). These time intervals are illustrated in Figure \ref{fig:segment}. We were interested in how these features changed during these intervals and how they are related to the interlocutors' behaviour during the exchange.

\subsection{Analysis results and discussion}

To compare the acoustic and visual features of the speaker and the exchange initiator in different types of conversational exchanges, we first compute the average value of each feature for the selected time intervals. For the speaker, we use the time interval from t1 to t2, which corresponds to the duration of their speech. For the exchange initiator, we use the time interval from t3 to t4, which corresponds to the moment when they start to take over the turn. By doing this, we can examine how the two roles differ in their use of prosody and facial expressions and how these features vary depending on the exchange type.

\begin{figure*}[!ht]
  \centering
  \includegraphics[width=1\linewidth]{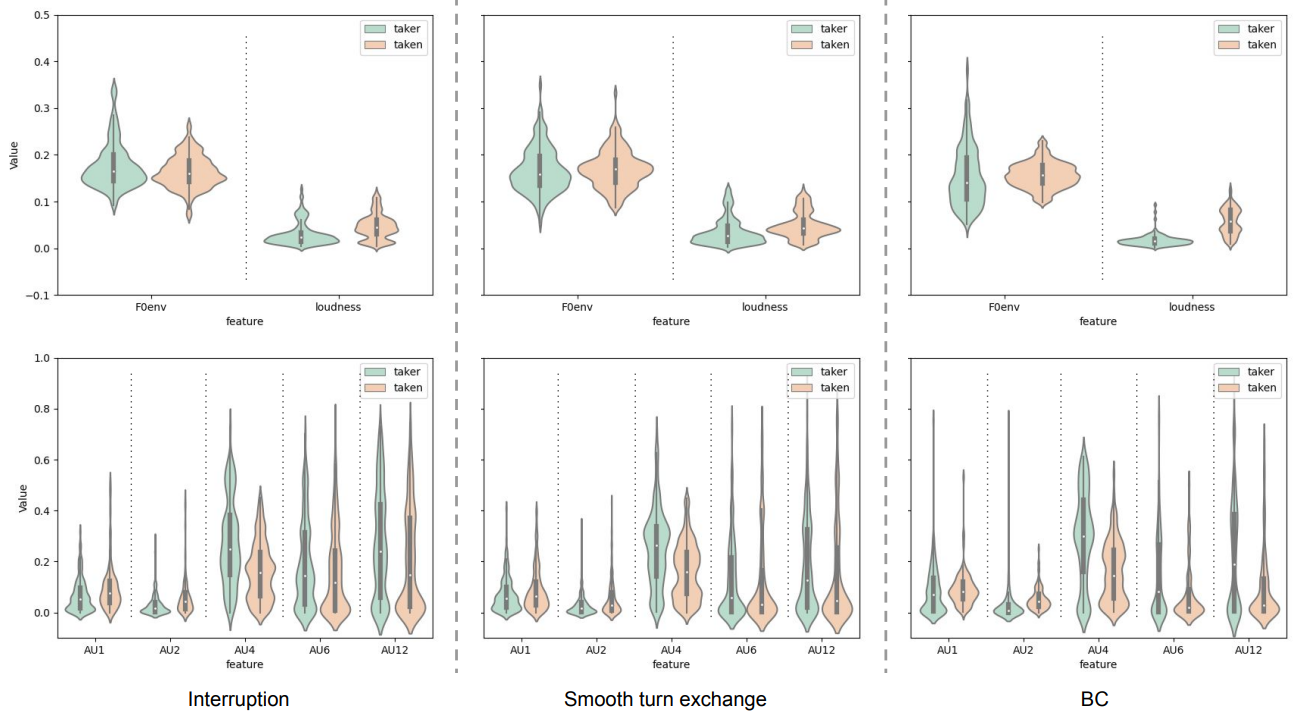}
  \caption{Average value of selected features during the corresponding intervals (speaker: t1 to t2, period taken; exchange initiator: t3 to t4, period taker).} 
  \label{fig:box_feat}
\end{figure*}

The distribution of the average value for each feature is shown in Figure \ref{fig:box_feat}, we perform a t-test (p-value< 0.05) to test for significant differences between the speaker and the exchange initiator for each feature and each exchange type. We find that the exchange initiator uses a higher pitch than the speaker when they initiate an interruption, but a lower loudness. This might suggest that they raise their voice frequency to signal their intention to interrupt, but do not increase their voice intensity to avoid being too aggressive. On the other hand, when they initiate a smooth turn exchange or a backchannel, they use a lower pitch and a lower loudness than the speaker. This indicates that they lower their voice frequency and intensity to show their agreement or acknowledgement and to smoothly transition to their turn. Moreover, we observe that the exchange initiator uses a higher pitch when they initiate an interruption than when they initiate a smooth turn exchange or a backchannel. This implies that they modulate their voice frequency according to the exchange type, using a higher pitch for more assertive exchanges and a lower pitch for more supportive ones. Similarly, we notice that the exchange initiator uses a lower loudness when they initiate a backchannel than when they initiate a smooth turn exchange or an interruption. 

For the visual features, we examine the facial action units (AUs) of the speaker and the exchange initiator, which represent the movements of different facial muscles. We find that the exchange initiator’s AU01 (inner brow raiser) and AU02 (outer brow raiser) have lower values than those of the speaker. On the other hand, the exchange initiator’s AU04 (brow lower), AU06 (cheek raiser) and AU12 (lip corner puller) have higher values than those of the speaker. In particular, when they initiate an interruption, compared to smooth turn exchange and backchannel, their AU06 and AU12 tend to be more active, which means that they smile more when they interrupt than when they cooperate or support. We also find that the speaker’s AU06 and AU12 are less active for the backchannel segment than for the other two types.

\begin{figure*}[!ht]
  \centering
  \includegraphics[width=1\linewidth]{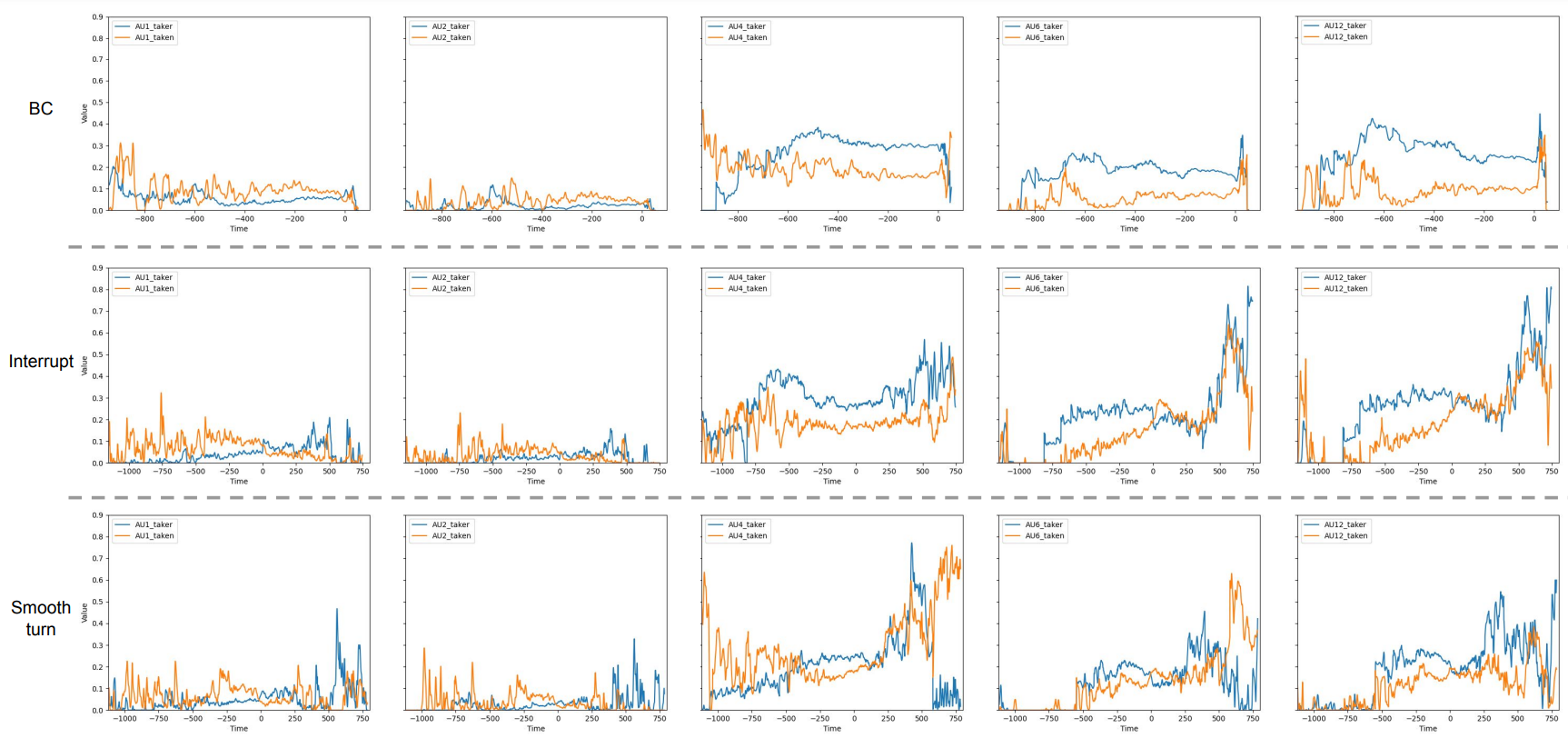}
  \caption{Average value over time (represented by frame in the figure, 25 frames per second) for each selected feature during the corresponding intervals (t1 to t4). } 
  \label{fig:curve}
\end{figure*}

\begin{figure*}[!ht]
  \centering
  \includegraphics[width=1\linewidth]{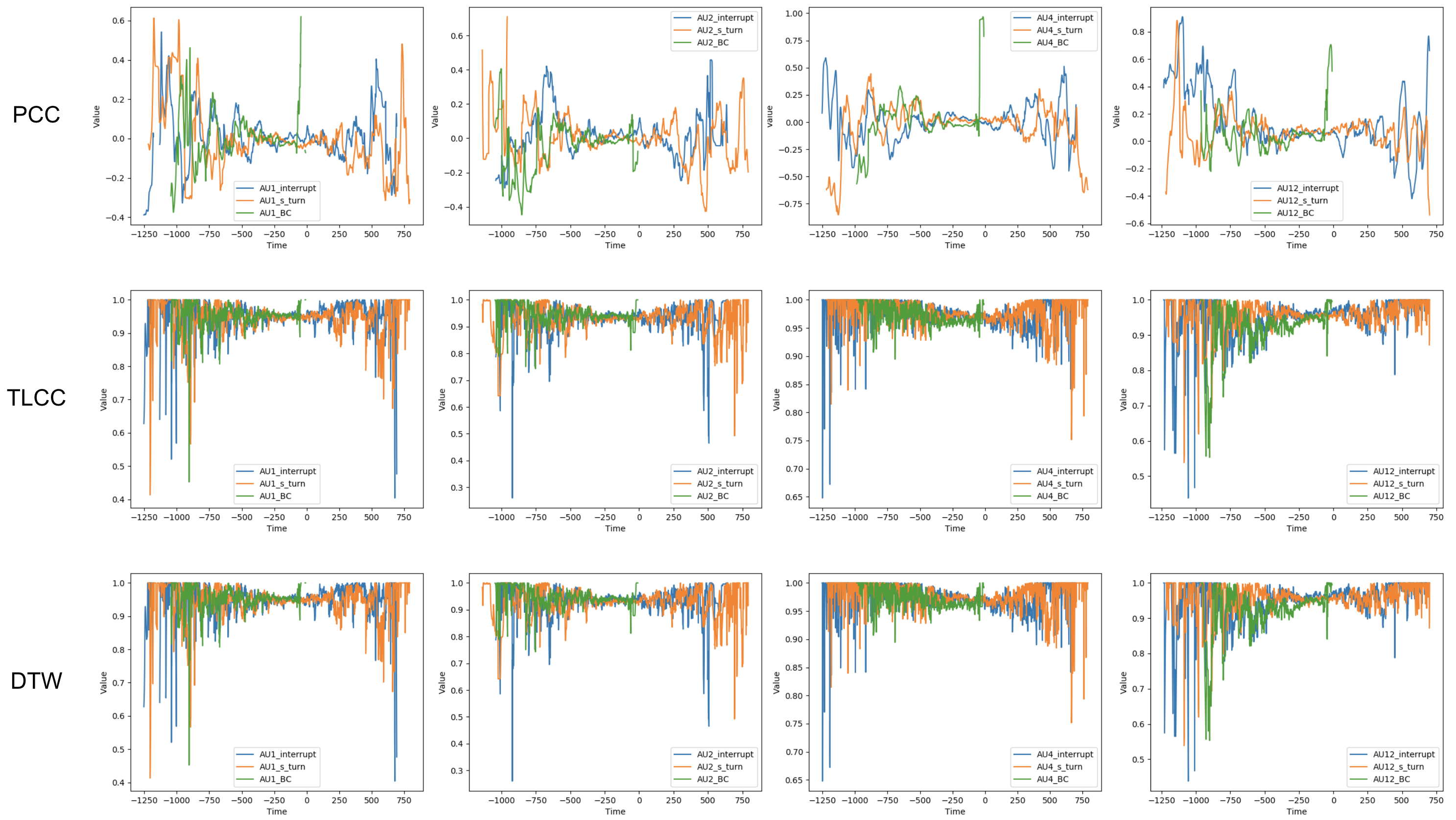}
  \caption{Average value over time (represented by frame in the figure, 25 frames per second) for each selected feature during the corresponding intervals (t1 to t4) for the synchrony features of PCC, TLCC and DTW.} 
  \label{fig:sync}
\end{figure*}

We then want to compare the evolution of visual features of the two roles (speaker and exchange initiator) and the different types of exchanges (smooth turn, interruption and backchannel) in a conversation. We select several visual features that are relevant to our analysis: AU01(inner brow raiser), AU02(outer brow raiser), AU04(brow lower), AU06(cheek raiser) and AU12(lip corner puller). All exchange samples are centred by the exchange onset point, which means for each sample, the timestamp t3 is always at 0 in the figure. We then calculate the mean value of each feature for each role and each exchange type over the time period t1 to t4 (see Figure \ref{fig:segment}). We plot these mean values as curves in Figure \ref{fig:curve} to see how they change over time. 

From the figure, we can see some interesting patterns. For backchannel, the interlocutors’ AU01 and AU02 always have lower values than other AUs during the exchanges. Interlocutors’ AU04, AU06, and AU12 stay at a relatively stable level most of the time, except for some sudden peaks or valleys. For instance, we can see a peak in AU06 and AU12 for both interlocutors when a backchannel occurs. We can also see a valley in AU04 for the exchange initiator at the backchannel period. 

Another interesting pattern is that the visual features of the two interlocutors tend to be harmonious or synchronized during an interruption. For example, we can see a significant increase of AU04, AU06, and AU12 for both interlocutors after an interruption point. This might indicate that the interlocutors share similar reactions after being interrupted or interrupting someone else. 

However, there are also some differences between the two roles in terms of their visual features. For example, we can see that the speaker’s AU06 and AU12 start to increase sometime before the interruption arrives. On the other hand, the exchange initiator’s AU04, AU06, and AU12 decrease at the end of the IPU (inter-pausal unit). And for smooth turn exchange, the speaker’s AU04 decreases to a lower level when approaching the end of the turn, then re-increases to a high level after the exchange.

% Sync
We look into the evolution of synchrony features of the different types of exchanges (smooth turn, interruption, and backchannel). For the exploration, we choose to investigate the synchrony for AU01(inner brow raiser), AU02(outer brow raiser), AU04(brow lower), AU06(cheek raiser) and AU12(lip corner puller). Like the analysis of visual features, the mean value of each synchrony feature for each exchange type is computed over the same time period t1 to t4, shown in Figure \ref{fig:segment}. The temporal change of the mean values is illustrated in Figure \ref{fig:sync}.

Via the evolution curves of the synchrony features, we can remark that similar synchrony levels are reported for all three exchange types until t3-125 (corresponding to $5s$ before the annotation point. This implies that we start to adapt and synchronise our behaviours (facial AUs) from $5$ seconds before any exchange of turn. At this moment, we can observe that there is a considerable change in synchrony levels for the backchannel. Via the measures of PCC, TLCC and DTW, we can note a significant increase in the synchrony level of AU01, AU02, and AU12. For AU04, a sudden decrease can be noticed. Backchannel is a way of showing engagement in a conversation that can be expressed through synchronous behaviours. Thus, the augmentation of correlation between the interlocutors is an expected observation supporting that backchannel is indeed a sign of synchrony.

The intriguing pattern that has been reported above for interruption, its visual features being harmonious or synchronized, can be seen again via the synchrony features between the interlocutors. After the interruption point, a notable rise in synchrony levels is seen for AU01, AU02, and AU04 which supports our prior finding. 

These patterns suggest that the visual features of the interlocutors are influenced by both their roles and their exchange types in a conversation. By analyzing these features, we can gain more insights into how people communicate and interact with each other in different situations.

\section{Conclusion}

Exchange management is essential for human-agent interaction, as it affects the conversational flow and the quality of the interaction. We aim to create an Embodied Conversational Agent (ECA) that can interact with humans in a natural and engaging way, by understanding how humans communicate and handle different types of exchanges in conversations. We analyzed the multimodal signals that humans use to initiate and respond to exchanges, such as smooth turns, interruptions, and backchannels.

We discovered that interruptions are more intrusive and have longer overlaps than smooth turns and backchannels, which are more collaborative and have shorter overlaps. We also noticed that the exchange initiator adjusts their voice pitch depending on the type of exchange, using a higher pitch for more assertive exchanges and a lower pitch for more friendly ones.

We also examined the facial expressions that humans use during and after the exchanges and found that some of them are more active and aligned, depending on the roles and the types of exchanges. These patterns indicate that the visual features of the interlocutors are influenced by both the role during exchanges and the type of the exchange.

We hope this research can give more insights into developing ECAs that can handle different types of exchanges in conversations. This research can be seen as an invitation to further exploration of multimodal signals during turns, such as body gestures and movements, which are also important for interaction.

\section*{Acknowledgments}

This work was performed as a part of ANR-JST-CREST TAPAS (ANR-19-JSTS-0001) and IA ANR-DFG-JST Panorama (ANR-20-IADJ-0008) project.

\bibliographystyle{unsrt}

\end{document}